\documentclass[prc,preprint,showpacs,preprintnumbers,amsmath,amssymb,superscriptaddress]{revtex4}

\usepackage{graphicx}% Include figure files
\usepackage{dcolumn}% Align table columns on decimal point
\usepackage{bm}% bold math

\begin{document}

\title{Effects of geometric constraints on the nuclear multifragmentation
       process}

\author{S.R.\ Souza}
\affiliation{Instituto de F\'\i sica, Universidade Federal do Rio de Janeiro\\
Cidade Universit\'aria, CP 68528, 21941-972, Rio de Janeiro, Brazil}
\affiliation{Instituto de F\'\i sica, Universidade Federal do Rio Grande do Sul\\
Av. Bento Gon\c calves 9500, CP 15051, 91501-970, Porto Alegre, Brazil}
\author{R.\ Donangelo}
\affiliation{Instituto de F\'\i sica, Universidade Federal do Rio de Janeiro\\
Cidade Universit\'aria, CP 68528, 21941-972, Rio de Janeiro, Brazil}
\author{W.G.\ Lynch}
\author{M.B.\ Tsang}
\affiliation{Department of Physics and Astronomy and National Superconducting
             Cyclotron Laboratory, Michigan State University,\\
East Lansing, Michigan 48824}

\date{\today}% It is always \today, today,
             %  but any date may be explicitly specified

\begin{abstract}
We include in statistical model calculations the facts that in the nuclear 
multifragmentation process the fragments are produced within a given volume 
and have a finite size.
The corrections associated with these constraints affect the partition modes 
and, as a consequence, other observables in the process.
In particular, we find that the favored fragmenting modes strongly suppress 
the collective flow energy, leading to much lower values compared to what is 
obtained from unconstrained calculations.
This leads, for a given total excitation energy, to a nontrivial correlation 
between the breakup temperature and the collective expansion velocity.
In particular we find that, under some conditions, the temperature of the
fragmenting system may increase as a function of this expansion velocity,
contrary to what it might be expected.
\end{abstract}

% PACS, the Physics and Astronomy
                         % Classification Scheme.
%\keywords{Suggested keywords}%Use showkeys class option if keyword
                              %display desired
\maketitle

\begin{section}{Introduction}
\label{sect:introduction}

The determination of the nuclear caloric curve is of great theoretical interest
since it may help to clarify the physics underlying the breakup of nuclear systems 
into many fragments, {\it i.e.}, nuclear multifragmentation.
For instance, there has been intensive debate on whether negative heat capacities 
should be observed in nuclear systems 
\cite{dagostino2000,elliot2000,chomaz2000,Dasgupta2003,gross1997,bondorf1985a,shlomo2004,smmIsobaric},
not to mention the fundamental question of whether there are any clear
signatures of a liquid-gas phase transition in nuclear multifragmentation
\cite{pichon2005,ma1999,noneq1,lgpt1,lgpt2}.
Experimental studies have proven to be essential in providing insight into the
main properties of the phenomenon (see \cite{reviewBetty} and references therein).
The difficulties in extracting the key quantities for the problem from experiments 
have been extensively discussed \cite{TsangExxT2001}.
As a consequence of these difficulties, conflicting experimental observations have been reported
\cite{cc1,cc2,cc3,cc4,cc5,cc6,cc7,cc8} 
and, therefore, it has not yet been possible to obtain a clear picture for the
nuclear multifragmentation process.

In spite of these uncertainties, many features have been clearly established,
such as the appearance of an appreciable collective radial expansion in central 
heavy-ion collisions 
\cite{flow1,flow2,flow3,flow4,flow5,flow6,flow7,flow8,flowMSU2,flowSchussler1992,flowSchussler1993}.
This is intuitively consistent with the results obtained by dynamical
approaches (see, for instance,
\cite{Bondorf1994,exoticSouzaNgo,flowMDSouzaDonangelo,flowSchussler1993,flowSchussler1992}), 
in which matter is strongly compressed during the first
violent stages of the collision and expands afterwards.

Although statistical models have turned out to be quite successful in explaining
many properties observed experimentally \cite{gross1997,bondorf1995}, the 
calculation of this radial flow lies beyond the scope of those statistical treatments.
Therefore, the radial flow energy is taken as an input parameter in these 
statistical calculations, where it is assumed that its main effect is to subtract 
the energy associated with the radial expansion from the thermal motion
(see, for example, \cite{flowSchussler1992,flowMSU1}).
This picture has been criticized by some authors
\cite{flowMSU2,DasGupta2001} since the non-zero relative velocity between
different regions of the system could prevent matter within a given region 
to coalesce at the breakup stage.
This effect has been quantitatively investigated in ref.\ \cite{DasGupta2001}
using the lattice gas model.

In this work, we incorporate, in the Statistical Multifragmentation Model
(SMM) \cite{smm1,smm2,smm3}, effects associated with the finite size of the
fragments in the radial expansion, by imposing the constraint that they must
lie entirely inside the breakup volume.
Although the corrections mentioned above and discussed in ref.\ \cite{DasGupta2001} 
should also be considered, they will not be addressed here.
In sect. \ref{sect:flow} we present these modifications to the standard radial flow 
calculations. Their inclusion in the SMM, together with a brief review of this model,
is performed in sect.\ \ref{sect:smm}.
The main results are presented in sect.\ \ref{sect:results} and conclusions
are drawn in sect.\ \ref{sect:conclusions}.

\end{section}

\begin{section}{Radial collective expansion}
\label{sect:flow}

We initially consider fragments as point particles originating from
the breakup of a source characterized by its mass and atomic numbers $A_0$ and
$Z_0$, temperature $T$, besides its spherical breakup volume $V=4\pi R^3/3$.
If one assumes that matter expands radially with velocity $u(r)$, at a
distance $r$ from the source's center, the probability that the 
energy of a fragment lies between $\varepsilon$ and $\varepsilon+d\varepsilon$ 
is given by \cite{bondorfEnergySpectrum}

\begin{eqnarray}
P(\varepsilon,r)d\varepsilon =& &\frac{1}{\sqrt{\pi TE^{(i)}_{\rm flow}(r)}}
\exp\left(-[\varepsilon+E^{(i)}_{\rm flow}(r)]/T\right)\nonumber\\
& & \sinh\left[\frac{2\sqrt{\varepsilon E^{(i)}_{\rm flow}(r)}}{T}\right]
d\varepsilon
\label{eq:per}
\end{eqnarray}

\noindent
where $E_{\rm flow}^{(i)}(r)\equiv\frac{1}{2}m_i u(r)^2$ is the radial expansion 
energy at $r$, $m_i=mA_i$, where $m$ denotes the nucleon mass, and $A_i$ stands 
for the mass number of the $i$-th fragment of a partition of the system into $M_f$ 
pieces.
The average kinetic energy of this fragment can be readily calculated from
the above equation,

\begin{equation}
E_i(r)=\int_0^\infty \varepsilon P(\varepsilon,r) d\varepsilon \,=\frac{3}{2}T+E^{(i)}_{\rm flow}(r)\;.
\label{eq:eflowr}
\end{equation}

\noindent
If we now take into consideration that the fragment has a finite size, and that 
it must lie, entirely, inside the breakup volume $V$, its average kinetic energy 
may be written as:

\begin{equation}
E_i= \int_0^{R-R_i} E_i(r) P_c(r) dr \;,
\label{eq:aveKi1}
\end{equation}

\noindent
where $R_i$ stands for the fragment's radius, and $P_c(r)$ is the probability 
that the fragment is created at a distance $r$ from the center.

If we assume that the expansion is irrotational and that the velocity field is given by

\begin{equation}
u(r)=\gamma \frac{r}{R}\;,
\label{eq:fieldvel}
\end{equation}

\noindent
where $\gamma$ is a constant, and also that the fragments may be formed with equal 
probability at any point inside the sphere of radius $R-R_i$, 
$P_c(r)=3r^2/(R-R_i)^3$, the average kinetic energy of the fragment is

\begin{equation}
E_i = \frac{3}{2}T+\frac{1}{2}mA_i\beta_{\rm flow}^2
\left[1-\frac{R_i}{R}\right]^2\;,
\label{eq:aveKi2}
\end{equation}

\noindent
where $\beta_{\rm flow}^2\equiv \frac{3}{5}\gamma^2$, thus clearly separating
the thermal motion and radial expansion contributions to the kinetic energy
of the fragment. 
Therefore, the total kinetic energy of the $M_f$ fragments of the partition is

\begin{equation}
E_{\rm trans}=\frac{3}{2}(M_f-1)T
+\varepsilon_{\rm flow} \sum_{A,Z} N_{A,Z}A\left[1-\frac{R_A}{R}\right]^2\;.
\label{eq:eKtotal}
\end{equation}

\noindent
One should notice that, following ref.\ \cite{smm2}, the center of mass motion
has been removed from the thermal contribution.
In the above expression, $N_{A,Z}$ denotes the multiplicity of a fragment
with mass and atomic numbers $A$ and $Z$, and we have defined
$\varepsilon_{\rm flow}\equiv\frac{1}{2}m\beta_{\rm flow}^2$.  
In the case where the geometric constraints are neglected, so that $R_A=0$ 
in the above expression, $\varepsilon_{\rm flow}$
represents the flow energy per particle, as the sum gives
$E_{\rm flow}=\varepsilon_{\rm flow}A_0$.
One sees that the inclusion of the finite size of the nuclear fragments 
clearly reduces the amount of energy in the radial expansion.
In particular, heavy fragments are more affected than light ones.
Therefore, since it influences the sharing between thermal and collective
energy in a way that depends on the fragment masses, this correction 
affects the partition modes, and, as a consequence, the values of other 
physical observables.

Finally, if we assume that the fragments are formed when the source has
expanded to $(1+\chi)$ of its volume at normal nuclear density, and that
the fragments when formed are at normal nuclear density, 
Eq.\ (\ref{eq:eKtotal}) can be rewritten as:

\begin{equation}
E_{\rm trans}=\frac{3}{2}(M_f-1)T+\sum_{A,Z}N_{A,Z}E^{\rm flow}_{A,Z}\;,
\label{eq:eKtotalf}\\
\end{equation}

\noindent
where 
\begin{equation}
E^{\rm flow}_{A,Z}=\varepsilon_{\rm flow} A 
\left[1-\left(\frac{A}{(1+\chi)A_0}\right)^{1/3}\right]^2\;.
\label{eq:eflowaz}
\end{equation}

In order to illustrate the magnitude of the corrections, we show, in
Fig.\ \ref{fig:factorvsa}, $E^{\rm flow}_{A,Z}/\varepsilon_{\rm flow}$
as a function of the mass number, for $A_0=168$, $\chi=2$, 5, and 9.
Comparison with the unconstrained results, {\it i.e.} 
$E^{\rm flow}_{A,Z}/\varepsilon_{\rm flow}=A$, shows that this effect is
important, even at very low densities.
Therefore, the predictions of the statistical calculations should be
modified when these constraints are included.

\begin{figure}[ht]
\includegraphics[angle=0,totalheight=7.0cm]{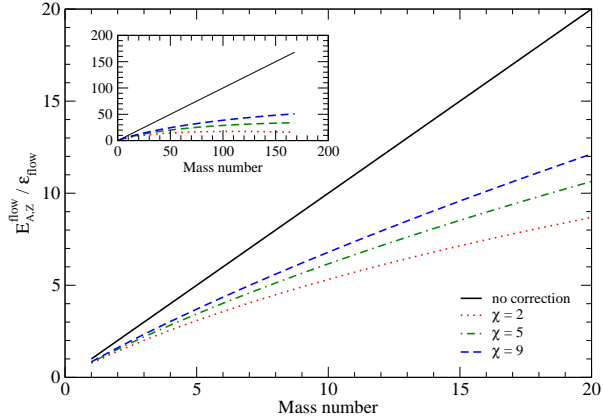}
\caption{\label{fig:factorvsa}
(Color online) $E^{\rm flow}_{A,Z}/\varepsilon_{\rm flow}$ as a function of
the mass number, for $A_0=168$.}
\end{figure}

\end{section}

\begin{section}{Inclusion into the Statistical Multifragmentation Model}
\label{sect:smm}

We briefly recall the main ingredients of the SMM. 
In it one assumes that the excited source undergoes a prompt statistical
breakup, subject to strict mass, charge, and energy conservation \cite{smm1,smm3,smm4}, 

\begin{equation}
A_0=\sum_{A,Z}N_{A,Z}A\;,\;\;\;Z_0=\sum_{A,Z}N_{A,Z}Z\;,
\label{eq:mcconst}
\end{equation}

\begin{equation}
E_0^{\rm g.s.}+E^*=\frac{3}{5}\frac{Z_0^2e^2}{R}+\sum_{AZ}N_{A,Z}E_{A,Z}(T,V)\;.
\label{eq:econst}
\end{equation}

\noindent
Above, $E_0^{\rm g.s.}$ represents the ground state energy of the source,
$E^*$ denotes the total excitation energy deposited into the system,
and $e$ is the elementary charge.
The fragment energies $E_{A,Z}$ have contributions from the translational motion, 
as well as from the nuclear bulk, surface, asymmetry, and Coulomb energies
\cite{smmIsobaric}.
The latter, is calculated through the Wigner-Seitz approximation
\cite{smm1,wignerSeitz}.
More specifically, $E_{A,Z}$ reads:

\begin{equation}
E_{A,Z}=-B_{A,Z}+E^*_{A,Z}+E^C_{A,Z}+\frac{3}{2}T+E^{\rm flow}_{A,Z}\;.
\label{eq:eaz}
\end{equation}

\noindent
We stress that the effects discussed in this work are contained in the
changes to the last term in the expression above, that were discussed in
the previous section. 
The binding energy of the fragments, $B_{A,Z}$, is calculated using the
prescription described in ref.\ \cite{smmMass}, whereas the remaining terms
read:

\begin{equation}
E^C_{A,Z}=-C_C\frac{Z^2}{A^{1/3}}\left(\frac{1}{1+\chi}\right)^{1/3}\;,
\label{eq:ecaz}
\end{equation}

\begin{equation}
E^*_{A,Z}=\frac{T^2}{\epsilon_0}A
+\left(\beta(T)-T\frac{d\beta}{dT}-\beta_0\right)A^{2/3}\;,
\label{eq:eex}
\end{equation}

\noindent
and

\begin{equation}
\beta(T)=\left[\frac{T^2_c-T^2}{T^2_c+T^2}\right]^{5/4}\;.
\label{eq:beta}
\end{equation}

\noindent
We take for all parameters the same values used  in ref.\ \cite{smmLong},
namely, a Coulomb parameter $C_C=0.720531$~MeV, bulk energy density
parameter $\epsilon_0=16.0$~MeV, critical temperature $T_c=18.0$~MeV, 
and  surface energy parameter $\beta_0=18.0$~MeV.
One should notice that by adding the term associated with the Coulomb
energy of the homogeneous sphere in Eq.\ (\ref{eq:econst}) to 
the Coulomb contributions given by the fragments' binding energies and
Eq.\ (\ref{eq:ecaz}), one obtains the Wigner-Seitz expression given
in ref.\ \cite{smm1}.
It is also worth mentioning that constraints on the center of mass
motion are also imposed for each breakup partition, so that the total kinetic
energy is given by Eq.\ (\ref{eq:eKtotalf}).

The breakup temperature is determined, for each partition, by solving
Eq.\ (\ref{eq:econst}), so that it is strongly dependent on the
partition mode. As the different terms in the sum appearing in Eq.\ (\ref{eq:econst})
are affected in different ways according to the size of the fragments they
represent, the temperature of the system will change appreciably from the
value calculated without geometrical constraints.

The average value of a physical observable $O_{A,Z}$ is
calculated through

\begin{equation}
\langle O_{A,Z}\rangle = \frac{\sum_fO_{A,Z}
\exp\left[\sum_{\{A,Z\}_f}N_{A,Z}S_{A,Z}\right]}
{\sum_f\exp\left[\sum_{\{A,Z\}_f}N_{A,Z}S_{A,Z}\right]}\;,
\label{eq:observable}
\end{equation}

\noindent
where the sum is performed over all possible partitions ${\{A,Z\}_f}$ of the nuclear
system into fragments, and the entropy of fragment $(A,Z)$, $S_{A,Z}$, is
calculated through the standard thermodynamic relation

\begin{equation}
S=-\frac{dF}{dT}
\label{eq:entropy}
\end{equation}

\noindent
where $F$ is the Helmholtz free energy. Since it depends on the
temperature of the fragmenting system, the weight of the corresponding
mode is also influenced by the constraints just described.
Ref.\ \cite{smmLong} provides a detailed presentation on how empirical
information is incorporated into $F$,
and we refer the reader to that work for details.
Except for the inclusion of the radial expansion, our SMM calculations follow
the description of the Improved Statistical Multifragmentation Model
(ISMM) presented in that work.

\begin{subsection}{Deexcitation of the primary fragments}
\label{sect:decay}
Since most excited fragments are detected after they have undergone
secondary decay, we have used the Weisskopf treatment described in
ref.\ \cite{BotvinaSD} to estimate these effects on the fragment energy
spectrum.
In this approach, the probability that a compound nucleus, with total
excitation energy $\varepsilon^*$, emits a fragment ($A$,$Z$), whose mass
is $\mu_{A,Z}$, is proportional to

\begin{equation}
\Gamma_{A,Z}(\varepsilon^*)=
\sum_{i=0}^{n}\int_0^{\varepsilon^*-b_{A,Z}-\varepsilon_{A,Z}^{(i)}}
f(\varepsilon)\,d\varepsilon\;,
\label{eq:Weisskopf1}
\end{equation}

\noindent
where

\begin{equation}
f(\varepsilon)=
\frac{g_{A,Z}^{(i)}\mu_{A,Z}\sigma_{A,Z}(\varepsilon)}{\pi^2\hbar^3}
\frac{\rho_{R}(\varepsilon^*-b_{A,Z}-\varepsilon_{A,Z}^{(i)}-\varepsilon)}
     {\rho_{CN}(\varepsilon^*)}\,\varepsilon\;.
\label{eq:Weisskopf2}
\end{equation}

\noindent
In the expression above, $b_{A,Z}$ represents the separation energy,
$g^{(i)}_{A,Z}$ denotes the spin degeneracy of the state $i$,
$\sigma_{A,Z}$ is the cross-section of the inverse reaction,
$\varepsilon_{A,Z}^{(i)}$ stands for the excitation energy of the emitted
fragment,
and $\rho(\varepsilon^*)$ corresponds to the density of states of either the
decaying nucleus (CN) or the residual fragment (R).
We have used the same parameters of ref.\ \cite{BotvinaSD}, except for
the binding energies and the level densities.
The former are the same used in our SMM calculations, whereas
the latter are given by the standard Fermi-gas expression
$\langle \varepsilon^*\rangle /A=aT^2$, but the excitation energy and the
breakup temperature
are taken as the average values obtained through Eq.\ (\ref{eq:observable})
for each primordial species.
Therefore, the density of states

\begin{equation}
\rho(\varepsilon^*)\propto \exp\left(2\sqrt{a\varepsilon^*}\right)
\label{eq:stateDensity}
\end{equation}

\noindent
has a different level density parameter $a$ for distinct primary fragment
species.
This ensures consistency with the population of the excited states in
SMM and in the secondary decay treatment.

The final kinetic energy spectrum is generated by a Monte Carlo sample
of the possible decay channels of the primary distribution.
More specifically, the excitation energy of a given primordial
fragment is selected with probability

\begin{equation}
P_E(\varepsilon^*)\propto\exp(-\varepsilon^*/T)\rho_{CN}(\varepsilon*)\;.
\label{eq:exstatpop}
\end{equation}

\noindent
The thermal velocity of the decaying fragment is then assigned according
to the Boltzmann distribution.

The radial expansion is incorporated by adding to the velocity a contribution
given by Eq.\ (\ref{eq:fieldvel}).
For consistency, the position of the fragment is uniformly sampled within a
spherical volume of radius $R$, which is equal to the breakup volume of
the system.
We also impose the constraint that the fragment must lie entirely inside it.
The contribution to the kinetic energy due to the Coulomb interaction is
estimated by considering the repulsion between the fragment and the remaining
part of the system.
We simply assume that the fragment with atomic number $Z_f$ is situated inside a
sphere of charge $(Z_0-Z_f)e$, homogeneously distributed within its volume.
The recoil of this core is taken into account when the corresponding boost
associated with this binary repulsion is added to the fragment's velocity.

The selection of a specific channel is made with probability

\begin{equation}
P_{A,Z}(\varepsilon^*)=\frac{\Gamma_{A,Z}(\varepsilon^*)}
{\sum_{\{A,Z\}}\Gamma_{A,Z}(\varepsilon^*)}\;,
\label{eq:channel}
\end{equation}

\noindent
where the sum runs over all possible decay channels.
We have considered the emission of all nuclei from $A=1$ to $A=10$.

For the selected deexcitation mode, the relative kinetic energy of the
products $\epsilon\le \varepsilon^*-b_{A,Z}-\varepsilon_{A,Z}^{(i)}$ is
sampled with weight proportional to Eq.\ (\ref{eq:Weisskopf2}).
Their velocities, in the rest frame of the decaying fragment, are determined
by energy and momentum conservation.
The excitation energy of the residue is then obtained by energy conservation
and it reads $\varepsilon^*_R=\varepsilon^*-b_{A,Z}-\varepsilon_{A,Z}^{(i)}
-\varepsilon$.
The decay chain is followed until the remnant fragment has a negligible
amount of excitation energy, {\it i.e.} it cannot decay by particle emition.

This Monte Carlo sample is repeated $10^5$ times for each primary species.
In the end, the multiplicities are weighed proportionally to the multiplicity
of the primordial fragments.

\end{subsection}

\end{section}

\begin{figure}[ht]
\includegraphics[angle=0,totalheight=7.0cm]{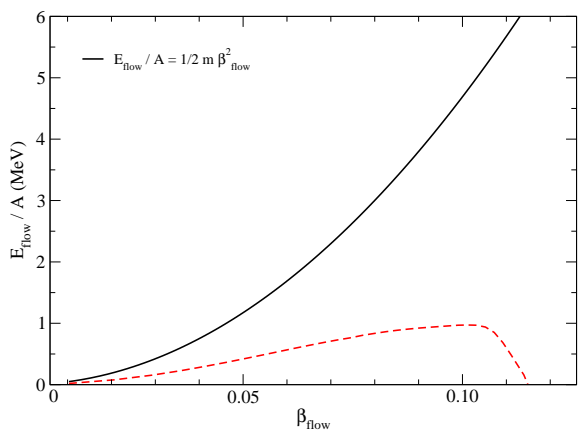}
\caption{\label{fig:efvsbeta}
(Color online) Average flow energy as a function of the average
flow velocity. The full curve illustrates the results without inclusion
of geometrical constraints whereas the dashed one corresponds to those with
their inclusion.
For additional details see the text.}
\end{figure}

\begin{section}{Results}
\label{sect:results}

To investigate the effects of the collective radial expansion in SMM, we
study the $A_0=168$ $Z_0=75$ system, at a fixed breakup density.
In Fig.\ \ref{fig:efvsbeta}, we show the average flow energy, $E_{\rm flow}$,
calculated through Eqs.\ (\ref{eq:eflowaz}) and (\ref{eq:observable}), as a 
function of the radial velocity $\beta_{\rm flow}$ (dashed line), in a case
where the system expanded to three times its volume at normal nuclear
density, {\it i.e.} $\chi=2$. 
The total available excitation energy of the system was taken to be
$E^*/A = 6$~MeV.
Comparison with the standard unconstrained values, represented in this picture
by the full line, demonstrates that the inclusion of the geometric constraints
dramatically suppresses the amount of energy which may be actually used in the 
radial collective expansion. 
One also observes that the flow energy reaches a maximum value, of approximately
$E_{\rm flow}/A = 1$~MeV at a value of $\beta_{\rm flow}=\beta_{\rm max}\approx 0.105$
close to the maximum possible (when all the energy available would appear
as radial flow), and then drops to zero as $\beta_{\rm flow}$ increases further.
This behaviour may be understood as a consequence of the fact that the total 
entropy of the system diminishes as more and more energy is stored into organized 
motion, reducing the accessible phase space associated with partitions leading
to large flow energy values.
Since the expansion velocity was taken to have a fixed value for all partitions, 
those which include large fragments, and consequently have smaller fragment 
multiplicities, are clearly favored, since they lead to smaller flow energies.

\begin{figure}[htb]
\includegraphics[angle=0,totalheight=7.0cm]{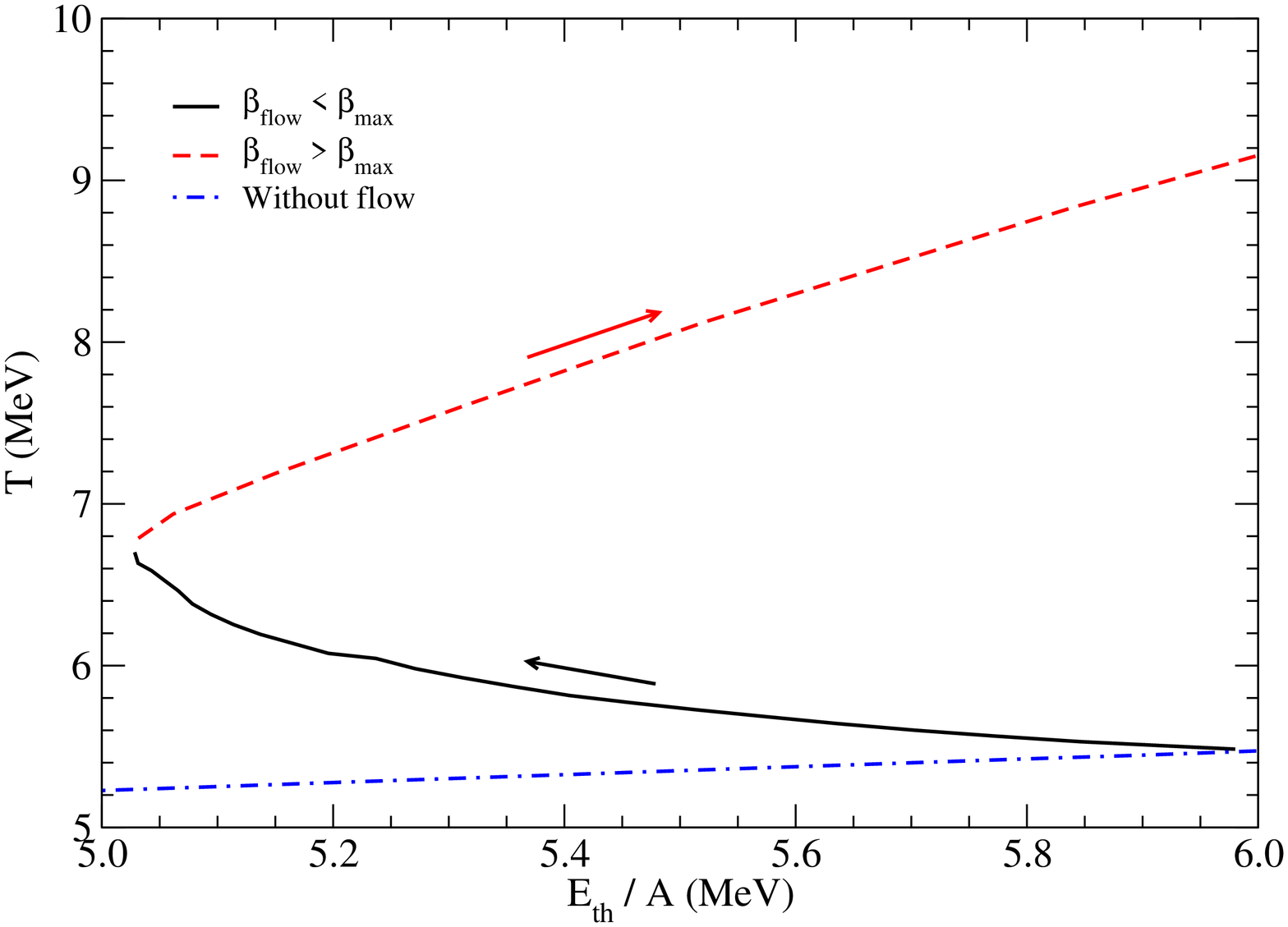}
\caption{\label{fig:tvseth}
(Color online) Average breakup temperature as a function of the average
thermal energy. The dash-dotted curve illustrates the results obtained
when geometrical constraints are not included. The full and dashed
curve are the results when these constraints are included: the full (dashed)
curve corresponds to velocity flow values below (above) the one leading to a 
maximum flow energy, as depicted in fig. \ref{fig:efvsbeta}. The arrows 
indicate the direction in which the radial velocity increases.}
\end{figure}

The changes on the preferred partitions reflect themselves on many observables,
such as the breakup temperature.
Indeed, SMM calculations at fixed breakup volume clearly show that the breakup
temperature becomes smaller if one simply removes the corresponding amount
of flow energy from the total excitation energy (see, for instance,
\cite{smmIsoscaling} and references therein).
This is illustrated by the dotted-dashed line in Fig.\ \ref{fig:tvseth}, which
displays the breakup temperature as a function of the thermal excitation energy,
in the case where geometrical constraints are disregarded.
On the other hand, the constraints associated with the collective motion
causes the breakup temperature, at a fixed total available excitation energy,
to rise instead of diminishing, as is also shown in this picture.
In this case, the thermal energy is defined as the difference between the total
available excitation energy and the average flow energy, {\it i.e.},
$E_{\rm th}=E^*-E_{\rm flow}$.
This behavior may be explained by the reduction of the fragment multiplicity,
which leads to larger fragments, moving with less flow energy.
The requirement of energy conservation, Eq.\ (\ref{eq:econst}), then leads to a 
higher temperature than when the constraints are not included.

\begin{figure}[t]
\includegraphics[angle=0,totalheight=7.0cm]{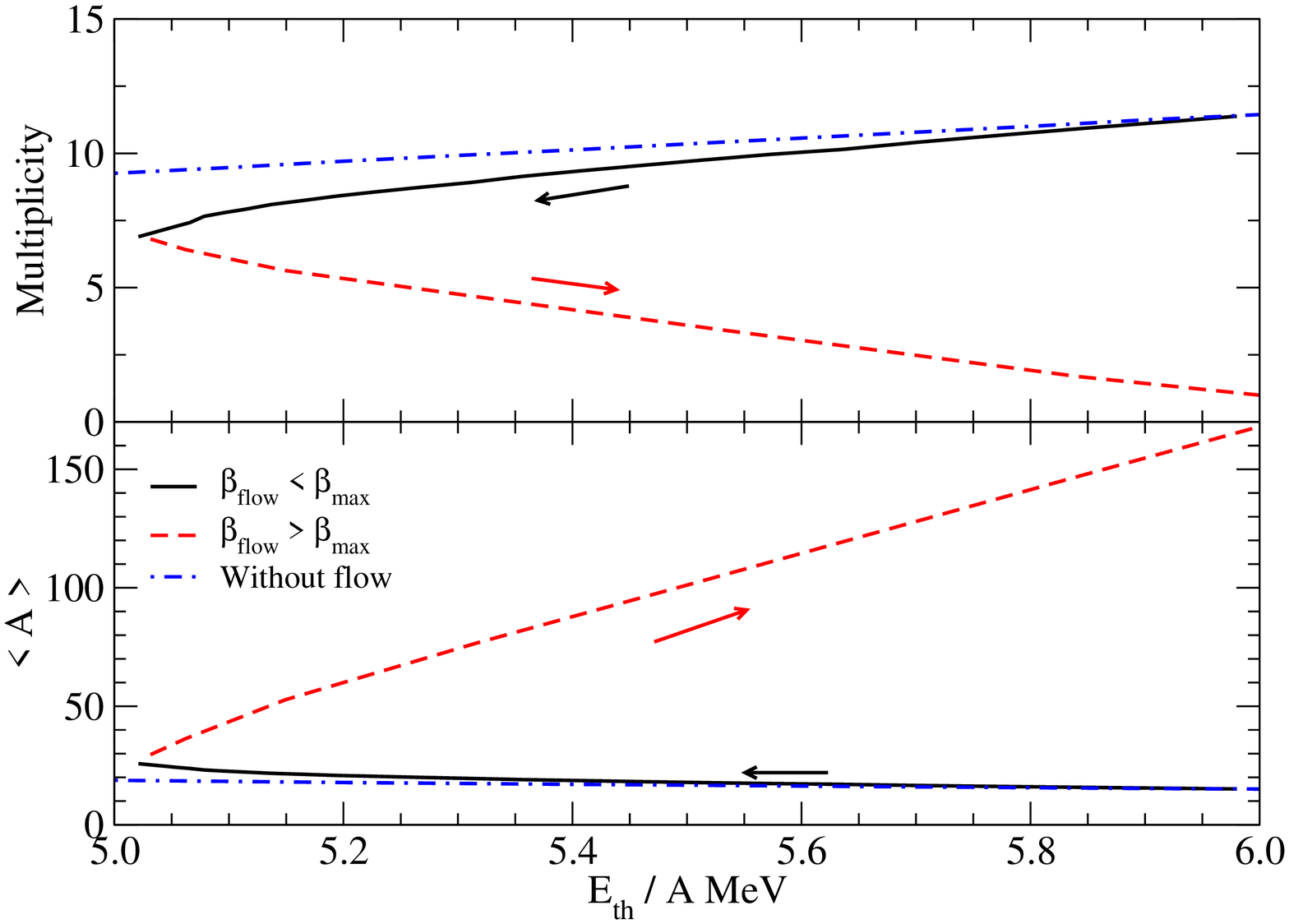}
\caption{\label{fig:maavvseth}
(Color online) Average primary multiplicity (upper panel) and average mass
number (lower panel) of fragments as a function of the thermal excitation
energy.
The curves correspond to the same cases illustrated in fig.\ref{fig:tvseth}.}
\end{figure}

The changes on the primary fragment multiplicity and on their average fragment
size 
are shown in Fig.\ \ref{fig:maavvseth} as a function of the thermal excitation energy.
As in the previous plot, the dashed-dotted line represents the results obtained
without geometrical constraints.
The inclusion of these constraints causes the fragment multiplicity to drop down
as the thermal excitation decreases, 
before the average flow energy reaches its maximum value, {\it i.e.}, for
$\beta_{\rm flow} < \beta_{\rm max}$.
Then, for $\beta_{\rm flow} > \beta_{\rm max}$, it keeps going down while 
the thermal energy increases until it reaches the smallest possible value $M_f=1$.
As expected, the opposite trend is observed for the average fragment size.

In spite of the important changes on the observables, the energy spectrum
of the particles still exhibits a shape which is similar to that expected
without constraints associated with the finite fragment sizes.
Indeed, the circles in Fig.\ \ref{fig:etransvsz} represent the average
kinetic energy of the primary fragments versus their atomic numbers.
The simulation has been carried out for $E^*/A=6.0$~MeV,
$\beta_{\rm flow}=\sqrt{2\varepsilon_{\rm flow}/m}$, and
$\varepsilon_{\rm flow}=2.0$~MeV.
As shown in that figure, the results can be fitted by the linear function 
$E_Z = 12.3+1.4Z$ MeV.
If the energy spectrum were interpreted disregarding the geometric constraints,
and one wrote $E_Z=\frac{3}{2}T+\varepsilon_{\rm flow}2Z$, comparison
with the fit would lead to $T=8.2$~MeV and 
$\varepsilon_{\rm flow}=0.7$~MeV.
However, the simulation gives $T=5.9$~MeV and, as already mentioned,
the expansion velocity corresponds to $\varepsilon_{\rm flow}=2.0$~MeV.
Therefore, the neglect of geometric constraints may lead to important
uncertainties in the interpretation of the experimental observations.

\begin{figure}[ht]
\includegraphics[angle=0,totalheight=7.0cm]{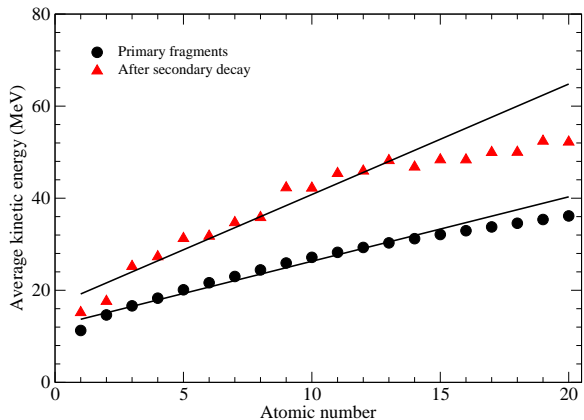}
\caption{\label{fig:etransvsz}
(Color online) Average frament kinetic energy as a function of the fragments'
atomic numbers. The lines correspond to a linear fit of the results.}
\end{figure}

In order to investigate the influence of the effects associated with the decay
of the hot primary fragments on the energy spectrum, we have employed
the deexcitation treatment presented in sect.\ \ref{sect:decay}.
The kinetic energy of the fragments after secondary decay is depicted
in Fig.\ \ref{fig:etransvsz} by the triangles.
As may be noticed, the slope of the spectrum increases appreciably and
one may adjust a linear function to reproduce its main trends.
Then, one finds $E_Z\approx 16.8+2.4Z$~MeV.
One sees that the Coulomb repulsion among the fragments appreciably
affects the slope of the distribution, besides the overall enhancement
of the fragment's kinetic energy.
Nevertheless, the slope is still much smaller than what would be given
by radial flow alone if the geometric constraints were not taken into account.
In spite of the great simplifications adopted in our deexcitation treatment,
we believe that the main effects are included in it, so that more refined
decay schemes should not change our conclusions significantly.
Therefore, our results suggest that the consistent treatment of the
geometrical constraints are very important in interpreting the
experimental observations.

\end{section}

\begin{section}{Concluding remarks}
\label{sect:conclusions}
We have investigated the effects of the inclusion of geometric constraints due
to the finite size of fragments in multifragmentation at a fixed breakup volume.
Our results show that the inclusion of these constraints in SMM lead to
qualitative different conclusions on the behavior of many physical observables
as the system undergoes a radial collective expansion.
In particular, our results suggest that radial flow alone should not be able
to explain a very large increase of the fragments kinetic energy as a function
of the atomic number.
Indeed, the simulations presented here show that, for a fixed total excitation
energy, the amount of energy stored in the radial expansion is strongly
suppressed.
As a consequence, other mechanisms
should be considered in order to explain the slopes
observed experimentally in the energy spectra of the fragments.
Thus, we believe that interpretations based on statistical calculations
in which energy flow is simply removed from the total energy should be
reviewed.

\end{section}

\begin{acknowledgments}
We would like to acknowledge CNPq, FAPERJ, and the PRONEX
program under contract No E-26/171.528/2006,
for partial financial support.
This work was supported in part by the National Science Foundation under Grant
No.\ PHY-01-10253 and INT-9908727.
\end{acknowledgments}

\end{document}